\documentclass[aps,prd,twocolumn,nofootinbib]{revtex4-2}
\usepackage[colorlinks]{hyperref}
\usepackage{amsmath,mathtools}
\usepackage{epsfig}
\usepackage{graphicx}
\usepackage{dcolumn}
\usepackage{amssymb,amsmath}
\usepackage{mathrsfs}
\usepackage{epsfig,bm}
\usepackage{bm}
\usepackage{booktabs}
\usepackage{siunitx}
\begin{document}

\title{Parity Nonconservation in Hydrogen Induced by Low-Mass Vector-Boson Exchange}

%\author{D. Budker }\email{budker@uni-mainz.de}
\author{V. A. Dzuba}\email{v.dzuba@unsw.edu.au}
\author{V. V. Flambaum}\email{v.flambaum@unsw.edu.au}
\author{G. K. Vong}\email{g.vong@unsw.edu.au}

\affiliation{School of Physics, University of New South Wales, Sydney 2052, Australia}

\begin{abstract}
Parity-nonconserving (PNC) effects in atoms produced by $Z$-boson exchange between the electron and the nucleus grow rapidly with the nuclear charge $Z$. If a hypothetical additional $Z'$ boson is light, however, its contribution does not exhibit the same strong enhancement with $Z$. As a result, the ratio of the low-mass $Z'$ contribution to the Standard Model $Z$-boson contribution increases rapidly with decreasing $Z$, in fact faster than $1/Z^2$. Hydrogen has a further important advantage: its theoretical description is substantially cleaner than that of heavy atoms, allowing a more accurate interpretation of experimental results. For these two reasons, hydrogen and deuterium  PNC experiments may provide an especially favorable setting in which to disentangle a possible $Z'$ contribution from the Standard Model background. 
In this paper we calculate the ratio of the $Z'$-boson contribution, for arbitrary $Z'$ mass, to the Standard Model $Z$-boson contribution to parity violation in hydrogen and deuterium, including both nuclear-spin-independent (NSI) and nuclear-spin-dependent (NSD) interactions.
\end{abstract}

\maketitle

\section{Introduction}
Parity nonconservation (PNC) in atoms provides one of the most precise low-energy tests of the electroweak interaction and a sensitive probe of new neutral-current physics  (see e.g. reviews
\cite{DF-PNC12,RevPNC}).
 %\cite{Khriplovich,Ginges2004,DK04,DF-PNC12}).
 The classic observation is that atomic transitions of opposite parity are weakly mixed by parity-violating electron--nucleus interactions, thereby generating otherwise forbidden electric-dipole amplitudes. Because these effects can be calculated in the Standard Model and measured with high precision, atomic PNC has long been used to test electroweak theory and to constrain extensions of it, including extra neutral gauge bosons ~\cite{BouchiatFayet2005,DienerGodfreyTuran2012,DFS17}.

Hydrogen occupies a special place in this program. In contrast to heavy many-electron atoms, hydrogen is free from many-body electronic-structure uncertainties and is sensitive in a particularly transparent way to proton neutral-current couplings. The possibility of studying parity violation in hydrogen was recognized long ago by Cahn and Kane~\cite{CahnKane1977}. Later analyses developed the formalism in detail for hydrogen and deuterium, emphasized the extraction of the NSI weak coupling constant $C_{1p}$ and the NSD weak coupling  constant  $C_{2p}$, and clarified the experimental advantages of level crossings and metastable states~\cite{DunfordHolt2007,DunfordHolt2011}. More recently, laser-based and quantum-control strategies for hydrogen PNC have been revisited, reinforcing the case that hydrogen can provide a clean determination of the proton weak charge and complementary information on weak neutral currents~\cite{RasorYost2020,LiDereviankoElliott2024}.

At the same time, parity-violating electron scattering has developed into a powerful and conceptually clean probe of semileptonic neutral-current interactions. The proton weak charge is especially sensitive to physics beyond the Standard Model because its Standard Model value is accidentally small~\cite{ErlerKurylovRamseyMusolf2003}. The final Qweak result for parity-violating elastic $ep$ scattering is in agreement with the Standard Model and places strong constraints on new semileptonic parity-violating interactions~\cite{AndroicEtAl2018}. In the heavy-boson limit these constraints admit a model-independent effective-field-theory interpretation in terms of contact operators, with the review of Ref.~\cite{CarliniVanOersPittSmith2019} providing the most useful summary of present bounds. Ongoing and future parity-violating electron-scattering programs further extend the sensitivity to extra neutral gauge bosons over a broad mass range~\cite{DevRodejohannXuZhang2021,ThomasWangWilliams2022,ThomasWangWilliams2025}.

The case of an additional neutral boson $Z'$ is therefore of renewed interest. For a sufficiently heavy $Z'$, exchange of the boson reduces at low energy to a local four-fermion interaction. In atomic language, its contribution can then be absorbed into shifts of the conventional weak couplings or weak charge. This short-range regime underlies most model-independent discussions of heavy $Z'$ constraints from parity-violation experiments~\cite{DienerGodfreyTuran2012,CarliniVanOersPittSmith2019}. However, if the boson mass is small enough that its Compton wavelength becomes comparable to or larger than the atomic length scale, the interaction is no longer local. One must retain the full Yukawa propagator, and the atomic matrix element depends on the detailed spatial overlap of electronic  wave functions and the  nuclear Yukawa-type  PNC potential rather than only on contact densities.

This finite-range regime has been studied in  our papers  Refs.  ~\cite{DFS17,Sr},  where the parity-violating potentials generated by a vector boson of arbitrary mass have been  applied  to several heavy atoms and ions. This  work provides the general framework needed to interpolate continuously between the contact and long-range limits. Yet, to the best of our knowledge, the hydrogen case has not been treated explicitly for a $Z'$ boson of arbitrary mass, despite the fact that hydrogen is arguably the cleanest system in which to isolate the dependence on the boson mass and on the underlying axial-electron/vector-proton  and vector-electron/axial-proton coupling structures. An additional advantage of hydrogen is that   the ratio of the low-mass $Z'$ contribution to the Standard Model $Z$-boson contribution increases rapidly with decreasing $Z$,  faster than $1/Z^2$  ~\cite{DFS17}. 

The purpose of the present work is to fill this gap. We investigate parity violation in hydrogen induced by exchange of a $Z'$ boson of arbitrary mass, including both nuclear-spin-independent (NSI) and nuclear-spin-dependent (NSD) interactions.
Our aim is not merely to reinterpret the standard weak interaction in a different notation, but to consider  the hydrogen PNC effects  in the regime where the finite range of the new interaction is essential and 
% $m_{Z'} a_0 \lesssim 1$,
the interaction cannot be reduced to a redefinition of the Standard Model $Z$-boson constants. Hydrogen is then especially attractive because the required matrix elements can be derived analytically or semi-analytically with controlled accuracy, making the dependence on $m_{Z'}$ transparent.

The analysis is also timely from the phenomenological point of view. Heavy-atom PNC, parity-violating electron scattering, and hydrogen spectroscopy probe related but not identical combinations of couplings ~\cite{ErlerKurylovRamseyMusolf2003,AndroicEtAl2018,CarliniVanOersPittSmith2019}. 

We use atomic units $\hbar=e=m_e=1$, $c=1/\alpha$.  

%^. Heavy atoms offer strong enhancement but involve many-electron structure; elastic $ep$ scattering yields the  model-independent constraints in the heavy-contact regime; hydrogen provides a clean bound-state system directly tied to proton structure and electroweak couplings~\cite{ErlerKurylovRamseyMusolf2003,AndroicEtAl2018,CarliniVanOersPittSmith2019}. A dedicated hydrogen calculation for arbitrary $m_{Z'}$ therefore helps clarify how these methods complement one another across the full boson-mass range.

\section{Interaction Lagrangian for \texorpdfstring{$Z$}{Z} and \texorpdfstring{$Z'$}{Z'} bosons.}

The neutral-current $Z$ Lagrangian is %($\hbar=c=1$)
\begin{equation}
\mathcal L_Z = -\frac14 Z_{\mu\nu}Z^{\mu\nu} + \frac12 m_Z^2 c^2 Z_\mu Z^\mu - \frac{g}{2\cos\theta_W} Z_\mu J_Z^\mu,
%\mathcal L_Z = -\frac14 Z_{\mu\nu}Z^{\mu\nu} + \frac12 m_Z^2  Z_\mu Z^\mu - \frac{g}{2\cos\theta_W} Z_\mu J_Z^\mu,
\end{equation}
with
\begin{equation}
J_Z^\mu = \sum_f \bar f\gamma^\mu\bigl(g_f^V-g_f^A\gamma_5\bigr)f.
\end{equation}

\subsection{Nuclear-spin-independent PV interaction}

The parity-violating nuclear-spin-independent (NSI)  part of the interaction  comes from the electron axial current and the proton vector current.  For the SM $Z$-boson (without radiative corrections) 
\begin{equation}
 g_e^A=-\frac12,
 \qquad
 g_p^V=\frac12-2\sin^2\theta_W,
 \end{equation}
Proton weak charge is equal to 
\begin{equation}
 Q_W^p=-4g_e^A g_p^V = 1-4\sin^2\theta_W.
\end{equation}
Inclusion of radiative corrections gives $ Q_W^p = 0.0705$ 
% $\QWp{}_{\rm SM}=\QWsm$ 
\cite{SM}.  Corrections produced by the dispersive parity violating interaction may also be included into the definition of the weak charge giving $ Q_W^p = 0.0719$ \cite{Samsonov}.

Then we may write the NSI part of the  $Z$-exchange parity-violating Hamiltonian as
\begin{equation}
V_Z(r) = -\frac{G_F}{2\sqrt2\,c}\,Q_W^p\,m_Z^2 c^2\,\Phi(m_Z,r)\,\gamma_5,
%V_Z(r) = -\frac{G_F}{2\sqrt2}\,Q_W^p\,m_Z^2 \,\Phi(m_Z,r)\,\gamma_5,
\label{eq:Vzfull}
\end{equation}
where $G_F
 \approx 2.2225 \times 10^{-14}$ a.u.
  is the Fermi constant of the weak interaction,
  \begin{equation}
\Phi(m,r) =  \frac{1}{4\pi} \frac{e^{-\mu  r}}{r}.
\end{equation}
 In atomic units $\mu=mc=m/ \alpha m_e$.
  %   $\Phi_m(r)$ is the Yukawa Green function folded with the proton density and normalized so that
%\begin{equation}
%(\nabla^2-m^2c^2)\Phi(m,r) = -\rho_N(r),
%\quad
%\int_0^\infty 4\pi r^2\rho_N(r)\,dr =1.
%\end{equation}
%The boson mass $m$ enters through dimensionless parameter 
%\begin{equation}
%\mu \equiv m/ \alpha m_e=268.173\,\frac{m}{\mathrm{MeV}/c^2}.
%\end{equation}
%with $m$ measured in electron-mass units when atomic units are used (in the natural units $\hbar=c=1$ we have 
%$\mu=m/ \alpha m_e$).  Numerically,
%\begin{equation}
%\mu = \frac{m_{Z'}/(\mathrm{MeV}/c^2)}{0.51099895}\,c \approx 268.173\,\frac{m_{Z'}}{\mathrm{MeV}/c^2}.
%\end{equation}
Finite nuclear radius must be introduced for a heavy atom case since the Dirac relativistic electron wave functions in $s_{1/2}$ and $p_{1/2}$ waves  tend to infinity for the point-like nucleus.  However, the non-relativistic approximation is sufficiently accurate for hydrogen,  so the  finite proton size correction is $\sim 10^{-5}$ and may be neglected. 
%In this case 
%\begin{equation}
%\Phi(m,r) =  \frac{1}{4\pi} \frac{e^{-\mu  r}}{r}
%\end{equation}
In the case of the large mediator mass, $m \to \infty$ , 
\begin{equation}
\Phi(m,r) =  \frac{1}{\mu^2} \delta^3(r)\,.
\end{equation}
%\begin{equation}
 %   V_{12}(r) = \frac{g_1^A g_2^V}{4\pi} \frac{e^{-m_{Z'} r}}{r} \gamma_5,
 %   \label{e:V12}
%\end{equation}

For a generic parity-violating $Z'$ boson we absorb the model-dependent gauge coupling and charges into two effective couplings $g_{eA}'$ and $g_{pV}'$ and present  NSI interaction potential as 
\begin{equation}
V_{Z'}(r) = g_{eA}' g_{pV}'\,\Phi(m_{Z'},r)\,\gamma_5 \,.
\label{eq:Vzprimepot}
\end{equation}

\subsection{ \texorpdfstring{$Z'$}{Z'} boson matrix elements}

We write the Dirac orbital as
\begin{equation}
\psi_{n\kappa m}(\mathbf r)
= \frac1r
\begin{pmatrix}
 g_{n\kappa}(r)\,\Omega_{\kappa m}(\hat{\mathbf r}) \\
 i f_{n\kappa}(r)\,\Omega_{-\kappa m}(\hat{\mathbf r})
\end{pmatrix},
\end{equation}
so that for $ns_{1/2}$ and $np_{1/2}$ states,
\begin{align}
    \mathcal M_n^{Z'} &\equiv \langle ns_{1/2}|V_{Z'}|np_{1/2}\rangle \nonumber \\ 
    &= i g_{eA}'g_{pV}'\int_0^\infty dr\,\bigl(g_{ns}f_{np}-f_{ns}g_{np}\bigr)\,\Phi_m(r).
    \label{eq:generalME}
\end{align}
In the nonrelativistic limit,
\begin{equation}
 g_{ns}(r)=rR_{ns}(r),
 \qquad
 g_{np}(r)=rR_{np}(r),
\end{equation}
where $R$ are the  non-relativistic hydrogen radial wave functions, while the small components follow from the Dirac radial equations,
\begin{align}
    f_{ns}(r)&=\frac{1}{2c}\left(\frac{d}{dr}-\frac1r\right)g_{ns}(r), \nonumber \\
 f_{np}(r)&=\frac{1}{2c}\left(\frac{d}{dr}+\frac1r\right)g_{np}(r).
\label{eq:smallcomponents} 
\end{align}

PNC effects in the transitions $1s-2s$, $2s-2p_{1/2}$,  $2s-3s$ and  $2s-4s$ have been discussed in the literature. Therefore, we need   matrix elements of the weak interaction   between close levels $2s-2p_{1/2}$,  $3s-3p_{1/2}$ and  $4s-4p_{1/2}$.
Specific matrix elements may be experimentally separated by reducing interval between opposite parity levels $ns-np_{1/2}$ to the close to zero value by magnetic field.
% For a point proton,
%\begin{equation}
%\Phi_m(r)=\frac{e^{-\mu r}}{4\pi r},
%\end{equation}
The hydrogenic integrals give
\begin{widetext}
%\end{widetext}
\begin{align}
\mathcal M_2^{Z'}
&= i\,\frac{g_{eA}'g_{pV}'}{96\pi c}\,\sqrt3\,\frac{3\mu+1}{(\mu+1)^3},
\\[0.4em]
\mathcal M_3^{Z'}
&= i\,\frac{g_{eA}'g_{pV}'}{27\pi c}\,\sqrt2\,
\frac{81\mu^3+54\mu^2+30\mu+4}{(3\mu+2)^5},
\\[0.4em]
\mathcal M_4^{Z'} &= i\,\frac{g_{eA}'g_{pV}'}{3840\pi c}\,\sqrt{15}\,
\frac{960\mu^5+800\mu^4+640\mu^3 +192\mu^2+42\mu+3}{(2\mu+1)^7}.
\label{eq:pointMEexplicit}
\end{align}
\end{widetext}

%For the  finite-nuclear-size matrix element for the hydrogen matrix elements considered here, the uniform-sphere correction is tiny: the finite-size contact matrix elements differ from the point-nucleus values by only about $1.6\times10^{-5}$.

\subsection{Standard Model \texorpdfstring{$Z$}{Z} matrix element and matrix-element ratios}

Because $m_Z c a_0\gg 1$, Eq.~\eqref{eq:Vzfull} reduces to the familiar contact interaction,
\begin{equation}
V_Z^{\rm contact}(r) = -\frac{G_F}{2\sqrt2\,c}\,Q_{Wp}\,\rho_N(r)\,\gamma_5.
\label{eq:Vzcontact}
\end{equation}
For a point proton this gives
\begin{equation}
\mathcal M_n^Z
= i\,\frac{G_FQ_{Wp}}{4\pi\sqrt2\,c^2}\,\frac{\sqrt{n^2-1}}{n^4},
\label{eq:Mzpoint}
\end{equation}
up to the overall phase convention of the $np_{1/2}$ state.  Below we use only absolute values.

It is convenient to define the dimensionless ratio
\begin{equation}
\eta_n(m)
\equiv
\left|\frac{\mathcal M_n^{Z'}}{\mathcal M_n^Z}\right|\,\frac{1}{|g_{eA}'g_{pV}'|}.
\label{eq:defeta}
\end{equation}
For a point proton one finds directly from Eqs.~\eqref{eq:pointMEexplicit} and \eqref{eq:Mzpoint},
\begin{widetext}
\begin{align}
\eta_2^{\rm point}(\mu) &= \frac{2\sqrt2\,c}{3G_FQ_{Wp}}\,
                            \frac{3\mu+1}{(\mu+1)^3},\\
%\\[0.4em]
\eta_3^{\rm point}(\mu) &= \frac{6\sqrt2\,c}{G_FQ_{Wp}}\,
                            \frac{81\mu^3+54\mu^2+30\mu+4}{(3\mu+2)^5},\\
%\\[0.4em]
\eta_4^{\rm point}(\mu) &= \frac{4\sqrt2\,c}{15G_FQ_{Wp}}\,\frac{960\mu^5+800\mu^4+640\mu^3 \mathstrut{+192\mu^2+42\mu+3}}{(2\mu+1)^7}.
\label{eq:pointetas}
\end{align}
\end{widetext}
The low-mass limits are therefore
\begin{align}
\left|\frac{\mathcal M_2^{Z'}}{\mathcal M_2^Z}\right|
&\xrightarrow[\mu\to 0]{}
|g_{eA}'g_{pV}'|\,\frac{\sqrt2\,c}{G_FQ_{Wp}}\,\frac23,
\\
\left|\frac{\mathcal M_3^{Z'}}{\mathcal M_3^Z}\right|
&\xrightarrow[\mu\to 0]{}
|g_{eA}'g_{pV}'|\,\frac{\sqrt2\,c}{G_FQ_{Wp}}\,\frac34,
\\
\left|\frac{\mathcal M_4^{Z'}}{\mathcal M_4^Z}\right|
&\xrightarrow[\mu\to 0]{}
|g_{eA}'g_{pV}'|\,\frac{\sqrt2\,c}{G_FQ_{Wp}}\,\frac45.
\label{eq:lowmasslimits}
\end{align}

In the high-mass limit, the Yukawa interaction also becomes contact-like, and
\begin{align}
\left|\frac{\mathcal M_n^{Z'}}{\mathcal M_n^Z}\right| 
\xrightarrow[\mu\to\infty]{}\,
& |g_{eA}'g_{pV}'|\,\frac{2\sqrt2\,c}{G_FQ_{Wp}\,\mu^2} ,
%\nonumber \\ =\, &|g_{eA}'g_{pV}'|\,\frac{2\sqrt2}{G_FQ_{Wp}\,c\,m^2}.
\label{eq:highmasslimit}
\end{align}
independently of $n$.
Table~\ref{tab:point} lists $\eta_n$ for a point proton.
\begin{table}[t]
\centering
\caption{Values of $\eta_n=|\mathcal M_n^{Z^\prime}/\mathcal M_n^{Z}|/|g_{eA}^\prime g_{pV}^\prime|$, using 
%$\QWp{}_{\rm SM}=0.0705$ 
$ Q_W^p = 0.0719$ 
\cite{SM}.}
\label{tab:point}
\small
\begin{tabular}{cccc}
\toprule
$m_{Z'}$ (MeV/$c^2$) & $n=2$ & $n=3$ & $n=4$ \\
\midrule
$10^{4}$ & \num{3.37e+04} & \num{3.37e+04} & \num{3.37e+04} \\
$10^{3}$ & \num{3.37e+06} & \num{3.37e+06} & \num{3.37e+06} \\
$10^{2}$ & \num{3.37e+08} & \num{3.37e+08} & \num{3.37e+08} \\
$10^{1}$ & \num{3.37e+10} & \num{3.37e+10} & \num{3.37e+10} \\
$10^{0}$ & \num{3.34e+12} & \num{3.34e+12} & \num{3.34e+12} \\
$10^{-1}$ & \num{3.06e+14} & \num{3.06e+14} & \num{3.06e+14} \\
$10^{-2}$ & \num{1.47e+16} & \num{1.45e+16} & \num{1.44e+16} \\
$10^{-3}$ & \num{7.15e+16} & \num{7.34e+16} & \num{7.34e+16} \\
$10^{-4}$ & \num{8.07e+16} & \num{9.05e+16} & \num{9.60e+16} \\
$10^{-5}$ & \num{8.08e+16} & \num{9.10e+16} & \num{9.70e+16} \\
\bottomrule
\end{tabular}
\end{table}

\subsection{Nuclear-spin-dependent PV interaction}

The Standard-Model Hamiltonian describing the nuclear-spin-dependent (NSD) parity-violating electron-proton interaction may be written as
\begin{equation}
H_{\rm SD} = \frac{G_F}{\sqrt{2}c} \frac{\varkappa_p}{I} \, {\bm \alpha}\cdot{\bm I} \, m_Z^2 c^2\,\Phi(m_Z,r),
\label{eq:HSDfull}
\end{equation}
where $\varkappa_p=0.043$ is the dimensionless NSD coupling constant \cite{ DunfordHolt2007}.
%For a proton in the Standard Model (without radiative corrections),
%\begin{equation}
%\varkappa_p = 4 g_e^V g_p^A = 1-4\sin^2\theta_W .
%\end{equation}
In the contact ($m_Z\to\infty$) approximation,
\begin{equation}
H_{\rm SD}^{\rm contact} = \frac{G_F}{\sqrt{2}c} \frac{\varkappa_p}{I} \, {\bm \alpha}\cdot{\bm I} \, \rho({\bm r}).
\label{eq:HSDcontact}
\end{equation}
For a generic parity-violating $Z'$ boson we absorb the model-dependent gauge coupling and charges into two effective couplings $g_{eV}'$ and $g_{pA}'$ and write the NSD interaction potential as
\begin{equation}
V_{Z'}(r) = \frac{g_{eV}' g_{pA}'}{I} \, {\bm \alpha}\cdot{\bm I}\,\Phi(m_{Z'},r).
\label{eq:VzprimeNSD}
\end{equation}

For hydrogen we take the proton to be nonrelativistic and at rest, and consider the stretched hyperfine state with maximal projections $I_z=J_z=1/2$, so that $F=F_z=1$. Then $( {\bm \alpha}\cdot{\bm I})/I \to \alpha_z$ and the matrix element reduces to
\begin{align}
\mathcal M_n^{Z',\rm SD}
\equiv
&\langle ns_{1/2},m_j=\tfrac12|V_{Z'}|np_{1/2},m_j=\tfrac12\rangle \nonumber \\
= & g_{eV}' g_{pA}'\!\int d^3r\,\psi_{ns}^\dagger(\mathbf r)\,\alpha_z\,\psi_{np}(\mathbf r)\,\Phi(m_{Z'},r).
\end{align}
Using the Dirac spinor convention of Eq.~(10) and the same nonrelativistic relations for the small components as in Eq.~\eqref{eq:smallcomponents}, one finds
\begin{equation}
\mathcal M_n^{Z',\rm SD}
= i g_{eV}'g_{pA}'\int_0^\infty dr\,\left(-\frac13 g_{np}f_{ns}-g_{ns}f_{np}\right)\Phi_m(r).
\label{eq:generalMEsd}
\end{equation}
%The factor $-1/3$ comes from the angular matrix element of $\sigma_z$ between the upper spinor harmonics for the $np_{1/2},m_j=1/2$ state, while the corresponding lower-component angular factor is $1$.

Evaluation of the hydrogenic radial integrals gives
\begin{widetext}
\begin{align}
\mathcal M_2^{Z',\rm SD}
&= -i\,\frac{g_{eV}'g_{pA}'}{288\pi c}\,\sqrt3\frac{9\mu^2+4\mu+1}{(\mu+1)^4},
\\[0.4em]
\mathcal M_3^{Z',\rm SD}
&
= -i\,\frac{g_{eV}'g_{pA}'}{81\pi c}\,\sqrt2\,
\frac{729\mu^4+324\mu^3+342\mu^2+72\mu+8}{(3\mu+2)^6},
\\[0.4em]
\mathcal M_4^{Z',\rm SD}
&
= -i\,\frac{g_{eV}'g_{pA}'}{11520\pi c}\, \sqrt{15}\,
\frac{5760\mu^6+2560\mu^5 
+4000\mu^4+1024\mu^3 +396\mu^2+48\mu+3}{(2\mu+1)^8}.
\label{eq:pointMEexplicitSD}
\end{align}
\end{widetext}

For the Standard-Model $Z$ boson, $m_Z c\gg 1$ and Eq.~\eqref{eq:HSDfull} may be replaced by the contact Hamiltonian \eqref{eq:HSDcontact}. For a point proton this yields
\begin{align}
\mathcal M_2^{Z,\rm SD}
&= -i\,\frac{G_F\varkappa_p}{32\pi\sqrt2\,c^2}\,\sqrt3,
\\[0.4em]
\mathcal M_3^{Z,\rm SD}
&= -i\,\frac{G_F\varkappa_p}{81\pi\sqrt2\,c^2}\,\sqrt2,
\\[0.4em]
\mathcal M_4^{Z,\rm SD}
&= -i\,\frac{G_F\varkappa_p}{512\pi\sqrt2\,c^2}\,\sqrt{15}.
\label{eq:MzpointSD}
\end{align}
Again, below we use only absolute values.

It is convenient to define the dimensionless ratio
\begin{equation}
\eta_n^{\rm SD}(m)
\equiv
\left|\frac{\mathcal M_n^{Z',\rm SD}}{\mathcal M_n^{Z,\rm SD}}\right|\,\frac{1}{|g_{eV}'g_{pA}'|}.
\label{eq:defetasd}
\end{equation}
Using Eqs.~\eqref{eq:pointMEexplicitSD} and \eqref{eq:MzpointSD}, one finds
\begin{widetext}
\begin{align}
\eta_2^{\rm SD}(\mu)
&= \frac{\sqrt2 c}{9G_F\varkappa_p}\,\frac{9\mu^2+4\mu+1}{(\mu+1)^4},
\\[0.4em]
\eta_3^{\rm SD}(\mu)
&= \frac{\sqrt2 c }{G_F\varkappa_p}\,
\frac{729\mu^4+324\mu^3+342\mu^2+72\mu+8}{(3\mu+2)^6},
\\[0.4em]
\eta_4^{\rm SD}(\mu)
&= \frac{2\sqrt2 c}{45G_F\varkappa_p}\,
\frac{5760\mu^6+2560\mu^5+4000\mu^4 +1024\mu^3+396\mu^2+48\mu+3}{(2\mu+1)^8}.
\label{eq:pointetasSD}
\end{align}
\end{widetext}

In the heavy-$Z$ and heavy-$Z'$ limit, all three ratios reduce to the same contact form, independently of $n$:
\begin{align}
\left|\frac{\mathcal M_n^{Z',\rm SD}}{\mathcal M_n^{Z,\rm SD}}\right|
\xrightarrow[\mu\to\infty]{} \,
&|g_{eV}'g_{pA}'|\,\frac{\sqrt2 c}{G_F\varkappa_p\,\mu^2} .
%\nonumber \\= \, &|g_{eV}'g_{pA}'|\,\frac{\sqrt2}{G_F\varkappa_p\,c m^2},
\label{eq:highmasslimitSD}
\end{align}

For a heavy Standard-Model $Z$ boson and a light $Z'$, the low-mass limits are
\begin{align}
\left|\frac{\mathcal M_2^{Z',\rm SD}}{\mathcal M_2^{Z,\rm SD}}\right|
&\xrightarrow[\mu\to 0]{}
|g_{eV}'g_{pA}'|\,\frac{\sqrt2c}{9G_F\varkappa_p},
\\
\left|\frac{\mathcal M_3^{Z',\rm SD}}{\mathcal M_3^{Z,\rm SD}}\right|
&\xrightarrow[\mu\to 0]{}
|g_{eV}'g_{pA}'|\,\frac{\sqrt2c}{8G_F\varkappa_p},
\\
\left|\frac{\mathcal M_4^{Z',\rm SD}}{\mathcal M_4^{Z,\rm SD}}\right|
&\xrightarrow[\mu\to 0]{}
|g_{eV}'g_{pA}'|\,\frac{2\sqrt2c}{15G_F\varkappa_p}.
\label{eq:lowmasslimitsSD}
\end{align}

Table~\ref{tab:pointSD} lists $\eta_n^{\rm SD}$ for a  point-like proton. As a test, we have also performed calculations for the NSI and NSD interactions using numerical solutions of the relativistic Dirac equations with a finite size proton, and confirmed  the non-relativistic analytical results for the point proton with the accuracy $\sim0.1\%$ . 
\begin{table}[t]
\centering
\caption{Values of $\eta_n^{\rm SD}=|\mathcal M_n^{Z^\prime,\rm SD}/\mathcal M_n^{Z,\rm SD}|/|g_{eV}^\prime g_{pA}^\prime|$, using $\varkappa_p^{\rm SM}=0.043$. } 
\label{tab:pointSD}
\vspace{1mm}
\small
\begin{tabular}{cccc}
\toprule
$m_{Z'}$ (MeV/$c^2$) & $n=2$ & $n=3$ & $n=4$ \\
\midrule
$10^{4}$ & \num{2.82e+04} & \num{2.82e+04} & \num{2.82e+04} \\
$10^{3}$ & \num{2.82e+06} & \num{2.82e+06} & \num{2.82e+06} \\
$10^{2}$ & \num{2.82e+08} & \num{2.82e+08} & \num{2.82e+08} \\
$10^{1}$ & \num{2.82e+10} & \num{2.82e+10} & \num{2.82e+10} \\
$10^{0}$ & \num{2.78e+12} & \num{2.78e+12} & \num{2.78e+12} \\
$10^{-1}$ & \num{2.48e+14} & \num{2.48e+14} & \num{2.48e+14} \\
$10^{-2}$ & \num{9.38e+15} & \num{9.20e+15} & \num{9.14e+15} \\
$10^{-3}$ & \num{2.37e+16} & \num{2.58e+16} & \num{2.63e+16} \\
$10^{-4}$ & \num{2.26e+16} & \num{2.55e+16} & \num{2.73e+16} \\
$10^{-5}$ & \num{2.25e+16} & \num{2.53e+16} & \num{2.70e+16} \\
\bottomrule
\end{tabular}
\end{table}

Results for deuterium PNC may be obtained by a simple rescaling of the hydrogen results. Deuterium PNC experiments are of independent interest since they may give us interaction constant of $Z'$ boson with neutron. Within the standard model,   deuterium weak charge $Q_W =-0.98207 N + 0.071918 Z=0.910$  is dominated by the neutron contribution and significantly exceeds the  proton weak charge. On the other hand, in the case of NSD interaction the standard model value  of $\varkappa_d$ is very small since $\varkappa_n \approx - \varkappa_p$. This leads to a large  relative enhancement of the NSD  $Z'$ boson contribution. A significant contribution is given by the nuclear anapole moment  \cite{anapole}.       
%\clearpage
\begin{acknowledgments}

This work was supported by the Australian Research Council Grants No. DP230101058.
\end{acknowledgments}

%\appendix 

%\bibliographystyle{apsrev4-2}
%\bibliography{dzuba,bibHPV}


\end{document}